%% file: CosmAcc.tex
\documentclass[11pt]{article}

\usepackage{amsmath}
\usepackage{amssymb}
\usepackage{graphicx}
\usepackage{subfigure}

\newcommand{\calD}{\ensuremath{\mathcal{D}}}
\newcommand{\avg}[1]{\ensuremath{\langle{#1}\rangle_\calD}}

\newcommand{\aD}{\ensuremath{a_\calD}}
\newcommand{\aDdot}{\ensuremath{\dot a_\calD}}
\newcommand{\aDddot}{\ensuremath{\ddot a_\calD}}
\newcommand{\calI}{\ensuremath{\mathcal{I}}}
\newcommand{\calP}{\ensuremath{\mathcal{P}}}
\newcommand{\eplog}{\ensuremath{\epsilon\ln\epsilon}}

\begin{document}

\title{The Possibility of Cosmic Acceleration via Spatial Averaging in
Lema\^\i tre-Tolman-Bondi Models}
\author{{\bf Aseem Paranjape}\footnote{E-mail:
    aseem@tifr.res.in}~~{\bf and}~{\bf T. P. Singh} 
\footnote{E-mail: tpsingh@tifr.res.in},\newline\\
 Tata Institute of Fundamental Research,\\
 Homi Bhabha Road,\\ Mumbai 400005, INDIA.}
\date{\today}

\maketitle

\input{files/abstract}
\input{files/intro}
\label{intro}
\input{files/avg}

\label{avg}
\input{files/LTBsoln}

\label{LTB}
\input{files/LTBavg}
\label{LTBavg}
\input{files/LTBmarg}

\label{LTBmarg}
\input{files/LTBunbd}

\label{LTBunbd}
\input{files/regular}

\label{regular}
\input{files/latetime}

\label{latetime}
\input{files/accln}
\label{accln}
\input{files/anlytc}

\label{anlytc}
\input{files/discuss}
\label{discuss}

\input{files/refer}
\end{document}

%% file: files/abstract.tex
\begin{abstract}
\noindent We investigate the possible occurrence of a positive cosmic
acceleration in a spatially averaged, expanding,  unbound
Lema\^\i tre-Tolman-Bondi cosmology. By studying an
approximation in which the contribution of three-curvature dominates
over the matter density, we construct numerical models which exhibit
acceleration.  

\end{abstract}

%% file: files/intro.tex
\begin{section}{Introduction}

We live in an inhomogeneous Universe, whose exact and  complicated
dynamics is described by Einstein's equations. It is generally assumed
that when  the spatial  inhomogeneities  are  averaged  over,  the
resulting  Universe  is described  by  the  standard  Friedmann
equations  for  a  homogeneous  and isotropic cosmology.  However, as
is  known  \cite{ellis},   since  Einstein equations are  non-linear,
the  averaging  over  the  inhomogeneous  matter distribution will in
general not yield the solution  of  Einstein  equations which is
described by the Friedmann-Robertson-Walker (FRW) metric. There  will
be  corrections  to the FRW solution, which could be small or large,
and  which could  in principle  lead  to  observational  effects
indicating  a departure   from standard  FRW cosmology.

The possibility  that  the  observed  cosmic  acceleration
\cite{reiss}  is caused by the spatial averaging  of  the  observed
inhomogeneities,  rather than by  a dark energy, has been investigated
and debated in the  literature \cite{ras1,kolb,wald}.  A systematic
framework  has  been  developed  for describing the dynamics of a
modified  Friedmann  universe,  obtained  after spatial averaging
\cite{buch1}. It has been  suggested  that,  within  the framework
of standard cosmology with cold dark  matter  initial  conditions, an
explanation of the acceleration in terms of averaged  inhomogeneities
is unlikely to work \cite{buch2}. However, it is perhaps fair  to
say  that the matter cannot be treated as completely  closed,    and
further  studies are desirable \cite{buch10}.

The Lema\^\i tre-Tolman-Bondi  (LTB)  cosmology \cite{LTB},
being  an exact solution  of Einstein equations for inhomogeneous dust
matter,  provides  a  useful  toy model for investigating the possible
connection  between  acceleration  and averaging of   inhomogeneities.
Various  authors  have  examined  different aspects of the  model  in
this  regard.  The  redshift-luminosity  distance relation  in  an
LTB  model  and  its  possible  connection   with   cosmic
acceleration, or the lack of it, have been studied  by  Celeri\'{e}r
\cite{cel}, Alnes et al. \cite{alnes} and by Vanderveld et
al. \cite{van}, Sugiura et  al. \cite{sug}, Mustapha et
al. \cite{mus}, Iguchi et al. \cite{igu}.  Nambu and  Tanimoto
\cite{nambu}  give  examples  of  cosmic  acceleration  after
averaging in an  LTB  model  consisting  of  a  contracting  region
and  an expanding region. Other works which study cosmic acceleration
in LTB  models are  those  by  Moffat  \cite{mof},   Mansouri
\cite{man}, Chuang et al. \cite{chu}, R\"{a}s\"{a}nen \cite{ras} and
Apostolopoulos et al. \cite{apos}.

It has sometimes been suggested in the literature that
both an expanding and a contracting region are needed for
acceleration. In the present  paper we will address a question which
does  not seem  to have been addressed in the above-mentioned works:
can spatial averaging  in a  universe  consisting  of  a
single   expanding LTB   region   produce acceleration? We show that
the answer is in the affirmative. We do  this  by considering a low
density, curvature dominated unbound LTB model in  which the
contribution  of  matter density  is  negligible   compared   to   the
contribution of the curvature  function.  Further, we concentrate on
the late time behaviour of such a model. As  a  result  of  this
proposed simplification, the calculation of the acceleration of  the
averaged scale factor becomes relatively simpler and conclusions about
acceleration can  be drawn, for specific choices of the energy
function.

In Section \ref{avg} of the paper  we  recall  the  effective  FRW
equations, resulting from spatial averaging in a dust dominated
spacetime.  In  Section \ref{LTB} we discuss spatial averaging for the
marginally bound LTB model and  point out there can be no acceleration
in this case.  The  unbound  LTB  model  is investigated  in  Section
\ref{LTBunbd},  in  the  approximation  that  the spatial  curvature
(equivalently, the energy function) dominates over the dust matter
density, and numerical and analytical examples of acceleration are
given.

\end{section}

%% file: files/avg.tex
\section{Averaging in Dust Dominated Spacetime}

For a general spacetime containing irrotational dust, the metric can
be written in synchronous and comoving gauge\footnote{Latin indices
  take values 1..3,  Greek indices take values 0..3. We set $c=1$.},
\begin{equation}
ds^2=\,-dt^2+h_{ij}(\vec{x},t)dx^idx^j\,.
\label{avg1}
\end{equation}
The expansion tensor $\Theta^i_j$ is given by $\Theta^i_j\equiv
(1/2)h^{ik}\dot h_{kj}$ where the dot refers to a derivative with
respect to  time $t$. The traceless symmetric shear tensor is defined
as  $\sigma^i_j\equiv \Theta^i_j-(\Theta/3) \delta^i_j$ where $\Theta
=\Theta^i_i$  is the expansion scalar. The Einstein equations can be
split  \cite{buch1} into a set of scalar equations and a set of vector
and traceless tensor equations. The scalar equations are the
Hamiltonian constraint \eqref{avg2a} and the evolution equation for
$\Theta$ \eqref{avg2b},
\begin{subequations}
\begin{equation}
^{(3)}\mathcal{R}+\frac{2}{3}\Theta^2-2\sigma^2=16\pi G\rho
\label{avg2a}
\end{equation}
\begin{equation}
^{(3)}\mathcal{R}+\dot\Theta+\Theta^2=12\pi G\rho
\label{avg2b}
\end{equation}
\end{subequations}
where the dot denotes derivative with respect to time $t$,
$^{(3)}\mathcal{R}$ is the Ricci scalar of the 3-dimensional hypersurface
of  constant $t$ and $\sigma^2$ is the rate of shear  defined by
$\sigma^2\equiv (1/2)\sigma^i_j\sigma^j_i$. Eqns. \eqref{avg2a} and
\eqref{avg2b} can be combined to give Raychaudhuri's equation
\begin{equation}
\dot\Theta+\frac{1}{3}\Theta^2+2\sigma^2+4\pi G\rho=0\,.
\label{avg3}
\end{equation}
The continuity equation $\dot\rho=-\Theta\rho$ which gives the
evolution of $\rho$, is consistent with Eqns. \eqref{avg2a},
\eqref{avg2b}. We only consider the scalar equations, since the
spatial average of a   scalar quantity can be defined in a gauge
covariant manner within a given  foliation of space-time.
For the space-time described by \eqref{avg1}, the spatial average of a scalar
$\Psi(t,\vec{x})$ over a {\em comoving}  domain \calD\ at time $t$ is
defined by
\begin{equation}
\avg{\Psi}=\frac{1}{V_\calD}\int_\calD{d^3x\sqrt{h}\,\Psi}
\label{avg4}
\end{equation}
where $h$ is the determinant of the 3-metric $h_{ij}$ and $V_\calD$ is
the volume of the comoving domain given by
$V_\calD=\int_\calD{d^3x\sqrt{h}}$.

Spatial averaging is, by definition, not generally covariant. Thus the
choice of foliation is relevant, and should be motivated on physical
grounds. In the context of cosmology, averaging over freely-falling
observers is a natural choice, especially when one intends to compare
the results with standard FRW cosmology. Following the definition
(\ref{avg4}) the following commutation relation then holds
\cite{buch1}  
\begin{equation}
\avg{\Psi}^\cdot-\avg{\dot\Psi}=
\avg{\Psi\Theta}-\avg{\Psi}\avg{\Theta}
\label{avg5}
\end{equation}
which yields for the expansion scalar $\Theta$
\begin{equation}
\avg{\Theta}^\cdot-\avg{\dot\Theta}=
\avg{\Theta^2}-\avg{\Theta}^2\,.
\label{avg6}
\end{equation}
Introducing the dimensionless scale factor
$\aD\equiv\left(V_\calD/V_{\calD in}\right)^{1/3}$ normalized by the
volume of the domain \calD\ at some initial time $t_{in}$, we can
average the scalar Einstein equations \eqref{avg2a}, \eqref{avg2b} and
the continuity  equation to obtain \cite{buch1}
\begin{subequations}
\begin{equation}
\avg{\Theta}=3\frac{\aDdot}{\aD}\,,
\label{avg7a}
\end{equation}
\begin{equation}
3\left(\frac{\aDdot}{\aD}\right)^2-8\pi G\avg{\rho}+
\frac{1}{2}\avg{\mathcal{R}}=\,-\frac{\mathcal{Q}_\calD}{2}\,,
\label{avg7b}
\end{equation}
\begin{equation}
3\left(\frac{\aDddot}{\aD}\right)+4\pi
G\avg{\rho}=\mathcal{Q}_\calD\,,
\label{avg7c}
\end{equation}
\begin{equation}
\avg{\rho}^\cdot=\,-\avg{\Theta}\avg{\rho}=\,-
3\frac{\aDdot}{\aD}\avg{\rho}\,.
\label{avg7d}
\end{equation}
\end{subequations}
Here $\avg{\mathcal{R}}$, the average of the spatial Ricci scalar
$^{(3)}\mathcal{R}$, is a domain dependent spatial constant. The
`backreaction' $\mathcal{Q}_\calD$ is given by
\begin{equation}
\mathcal{Q}_\calD\equiv\frac{2}{3}\left(\avg{\Theta^2}-
\avg{\Theta}^2\right)-2\avg{\sigma^2}
\label{avg8}
\end{equation}
and is also a spatial constant. The last equation \eqref{avg7d} simply
reflects the fact that the mass contained in a comoving domain is
constant by construction : the local continuity equation
$\dot\rho=-\Theta\rho$ can be solved to give
$\rho\sqrt{h}=\rho_0\sqrt{h_0}$ where the subscript $0$ refers to some
arbitrary reference time $t_0$. The mass $M_\calD$ contained in a
comoving domain \calD\ is then $M_\calD=\int_\calD{\rho\sqrt{h}d^3x}
=\int_\calD{\rho_0\sqrt{h_0}d^3x}=\,$constant. Hence
\begin{equation}
\avg{\rho}=\frac{M_\calD}{V_{\calD in} \aD^3}
\label{avg9}
\end{equation}
which is precisely what is implied by Eqn. \eqref{avg7d}.

This averaging procedure can only be applied for spatial scalars, and
hence only a subset of the Einstein equations can be smoothed out. As
a result it may appear that the outcome of such an approach is
severely restricted, and essentially incomplete due to the
impossibility to analyse the full set of equations. However one should
note that the cosmological parameters of interest are scalars, and the
averaging of the exact scalar part of Einstein equations provides the
requisite needed information. A more general strategy would be to
consider the smoothing of tensors, which is beyond the scalar approach
that certainly provides useful information, albeit not the full
information.

Equations \eqref{avg7b}, \eqref{avg7c} can be cast in a form which is
immediately comparable with the standard FRW equations
\cite{buch4}. Namely,
\begin{subequations}
\begin{equation}
\frac{\aDddot}{\aD}=\,-\frac{4\pi G}{3}\left(\rho_{\rm eff}+
3P_{\rm eff}\right)
\label{avg10a}
\end{equation}
\begin{equation}
\left(\frac{\aDdot}{\aD}\right)^2=\frac{8\pi G}{3}\rho_{\rm eff}
\label{avg10b}
\end{equation}
\end{subequations}
with $\rho_{\rm eff}$ and $P_{\rm eff}$ defined as
\begin{equation}
\rho_{\rm eff}=\avg{\rho}-\frac{\mathcal{Q}_\calD}{16\pi G}-
\frac{\avg{\mathcal{R}}}{16\pi G}~~~\text{;}~~~
P_{\rm eff}=\,-\frac{\mathcal{Q}_\calD}{16\pi
  G}+\frac{\avg{\mathcal{R}}}{48\pi G}\,.
\label{avg11}
\end{equation}
A necessary condition for \eqref{avg10a} to integrate to
\eqref{avg10b} takes the form of the following differential equation
involving $\mathcal{Q}_\calD$ and $\avg{\mathcal{R}}$
\begin{equation}
\mathcal{\dot Q}_\calD+6\frac{\aDdot}{\aD}\mathcal{Q}_\calD
+\avg{\mathcal{R}}^{\cdot}+2\frac{\aDdot}{\aD}\avg{\mathcal{R}}=0
\label{avg12}
\end{equation}
and the criterion to be met in order for the effective scale factor
$\aD$ to accelerate, is
\begin{equation}
\mathcal{Q}_\calD>4\pi G\avg{\rho}\,.
\label{avg13}
\end{equation}

%% file: files/LTBsoln.tex
\section{The LTB Solution}
The system of equations \eqref{avg10a}, \eqref{avg10b} and
\eqref{avg12} is only consistent, it does not close. For a completely
general spacetime with dust, therefore, it is not possible to proceed
with the analysis without making certain assumptions about the form of
the functions $\mathcal{Q}_\calD$ and $\avg{\mathcal{R}}$
\cite{buch1,buch2}. For this reason, it becomes convenient to work
with the LTB metric, an exact solution of the Einstein equations which
is a toy model consisting of a spherically symmetric inhomogeneous
dust dominated spacetime. In this section, we describe the LTB
solution and apply to it the averaging procedure described above for
the simplest, marginally bound case. In the next section we extend the
analysis to the unbound LTB solution. The LTB metric for pressureless
dust is given in the synchronous and comoving gauge, by 
\begin{equation}
ds^2=\,-dt^2+\frac{R^{\prime 2}(r,t)}{1+2E(r)}dr^2+
R^2(r,t)\left(d\theta^2 + \sin^2\theta d\phi^2\right)\,.
\label{LTB1}
\end{equation}
The Einstein equations simplify to
\begin{subequations}
\begin{equation}
\frac{1}{2}\dot R^2(r,t)-\frac{GM(r)}{R(r,t)}=E(r)\,,
\label{LTB2a}
\end{equation}
\begin{equation}
4\pi\rho(r,t)=\frac{M^\prime(r)}{R^\prime(r,t)R^2(r,t)}\,.
\label{LTB2b}
\end{equation}
\end{subequations}
Surfaces of constant $r$ are $2-$spheres having area $4\pi R^2(r,t)$. 
$\rho(r,t)$ is the energy density of dust, while $E(r)$ and $M(r)$ are
arbitrary functions that arise  on integrating the dynamical
equations. Solutions can be found for three  cases $E(r)>0$, $E(r)=0$
and $E(r)<0$. We will restrict our attention to models in which $E(r)$
has the same sign, for all $r$. The solution for $E(r)=0$  (the
marginally bound case) has the particularly simple form 
\begin{equation}
R(r,t)=\bigg(\frac{9GM(r)}{2}\bigg)^{1/3}
\left(t-t_0(r)\right)^{2/3},~~~~\text{for}~E(r)=0\,.
\label{LTB3}
\end{equation}
Here $t_0(r)$ is another arbitrary function arising from integration.
The solution describes an expanding region, with the initial time
$t_{in}$ chosen such that $t>t_{in}\geq t_0(r)$ for all $r$. For the
other two cases,  the solutions can be written in parametric form 
\begin{subequations}
\begin{equation}
R=\frac{GM(r)}{2E(r)}\left(\cosh\eta -1\right)~~~
;~~~t-t_0(r)=\frac{GM(r)}{\left(2E(r)\right)^{3/2}}\left(\sinh\eta
- \eta\right)~,~0\leq\eta<\infty~,~~~\text{for}~E(r)>0\,.
\label{LTB4a}
\end{equation}
\begin{equation}
R=\frac{GM(r)}{-2E(r)}\left(1-\cos\eta\right)~~~;~~~
t-t_0(r)=\frac{GM(r)}{\left(-2E(r)\right)^{3/2}}
\left(\eta -\sin\eta\right)~,~0\leq\eta\leq 2\pi,~~~
\text{for}~E(r)<0\,.
\label{LTB4b} 
\end{equation}
\end{subequations}
In the unbound case ($E(r)>0$), $R(r,t)$ increases monotonically with
$t$,  for every shell with label $r$. In the bound case ($E(r)<0$),
$R(r,t)$ increases to a maximum value $R_{max}(r)$ for each shell $r$
and then decreases back to $0$ in a finite time. 

In all cases, there are two physically different free functions,
although  three arbitrary functions $E$, $M$ and $t_0$ appear. One of
the three  represents the freedom to rescale the coordinate $r$. We
use this freedom to  set $R(r,t_{in})=r$. To completely specify the
solution, we   specify the initial density $\rho_{in}(r)$ and the
function $E(r)$. This specifies  $M(r)=4\pi\int_0^r{\rho_{in}(\tilde
r)\tilde r^2d\tilde r}$ (which in the marginally bound case is
interpreted as the mass contained  in a comoving shell), and $t_0(r)$
can be solved for using Eqns. \eqref{LTB3}, \eqref{LTB4a} or
\eqref{LTB4b} as the case may be, at time $t=t_{in}$.

%% file: files/LTBavg.tex
\subsection{Averaging the LTB Solution}
The quantities defined in Sec. \ref{avg} can be computed for the LTB
metric of Eqn. \eqref{LTB1}. The averages are computed over a spherical
domain of radius $r_\calD$, centered on the observer. Other choices of
the averaging domain will possibly yield different results, however,
the choice of a spherical domain seems natural for the spherically
symmetric metric of Eqn. \eqref{LTB1}. For clarity, we suppress the $r$
and $t$ dependences of the various functions in the following
\begin{equation}
V_\calD=4\pi\int_0^{r_\calD}{\frac{R^\prime
R^2}{\sqrt{1+2E}}~dr}~~\text{;}~~~
\avg{\Theta}=\frac{4\pi}{V_\calD}\int_0^{r_\calD}{\frac{R^2\dot
R^\prime+ 2R\dot RR^\prime}{\sqrt{1+2E}}~dr}=\frac{\dot
V_\calD}{V_\calD}~~
\text{;}~~~M_\calD=\int_0^{r_\calD}{\frac{M^\prime}{\sqrt{1+2E}}~dr} 
\label{LTBavg1}
\end{equation}
where a prime denotes a derivative with respect to $r$.
Note that $M_\calD\neq M(r_\calD)$ if $E\neq 0$. Only in the
marginally bound case is the function $M(r)$ identified with the mass
contained  in the shell with label $r$. It is convenient to work with
the combination  $(2/3)\avg{\Theta^2}-2\avg{\sigma^2}$ rather than
evaluate the average rate of shear $\avg{\sigma^2}$ separately. We
define this to be  $\mathcal{C}_\calD$ and obtain 
\begin{equation}
\mathcal{C}_\calD\equiv\frac{2}{3}\avg{\Theta^2}-2\avg{\sigma^2}=
\frac{8\pi}{V_\calD} \int_0^{r_\calD}{\frac{2R\dot R\dot R^\prime+\dot
R^2R^\prime}{\sqrt{1+2E}}~dr} \,.
\label{LTBavg2}
\end{equation}

%% file: files/LTBmarg.tex
\subsection{The marginally bound case - vanishing backreaction} 
The results of the previous subsection hold for all classes of the LTB
solution, provided the averaging domain is spherically symmetric about
the center. Now consider the marginally bound case $E(r)=0$ for all
$r$. The algebra in this case becomes very simple, and the
backreaction can be computed analytically. We will show next that for
a single domain with $E(r)=0$ throughout, the backreaction
$\mathcal{Q}_\calD$ is, in fact, zero. Also, the average spatial
curvature $\avg{\mathcal{R}}$ is zero (which is expected by inspection
of the metric \eqref{LTB1} if we note that a spatially uniform initial
density profile in the LTB solution in this case yields the
corresponding FRW solution). As described earlier we have  
\begin{equation}
M(r)=4\pi\int_0^r{\rho_{in}(\tilde r)\tilde r^2d\tilde
r}~~~~\text{;}~~~~
t_0(r)=t_{in}-\frac{r^{3/2}}{3}\sqrt{\frac{2}{GM(r)}}\,. 
\label{LTBmarg1} 
\end{equation}
To obtain the second equation we have used Eqn. \eqref{LTB3} at time
$t=t_{in}$ with the condition $R(r,t_{in})=r$. Some algebra then
yields the following  results
\begin{subequations}
\begin{equation}
\dot
R=\frac{2}{3}\left(\frac{9GM(r)}{2}\right)^{1/3}\left(t-t_0(r)\right)^{-1/3} 
\label{LTBmarg2a}
\end{equation}
\begin{equation}
R^\prime=\frac{2}{3}\left(\frac{9GM(r)}{2}\right)^{1/3} 
\left(t-t_0(r)\right)^{-1/3} 
\left[\frac{M^\prime(r)}{2M(r)}\left(t-t_0(r)\right)-t_0^\prime(r)\right]
\label{LTBmarg2b}
\end{equation}
\begin{equation}
\dot R^\prime=\frac{2}{9}\left(\frac{9GM(r)}{2}\right)^{1/3}
\left(t-t_0(r)\right)^{-4/3}
\left[\frac{M^\prime(r)}{M(r)}\left(t-t_0(r)\right)+t_0^\prime(r)\right]
\label{LTBmarg2c}
\end{equation}
\begin{equation}
V_\calD=6\pi GM_\calD\left(t-t_0(r_\calD)\right)^2~\Rightarrow~
\aD(t)=\left(\frac{t-t_0(r_\calD)}{t_{in}-t_0(r_\calD)}\right)^{2/3}
\label{LTBmarg2d}
\end{equation}
\begin{equation}
\avg{\Theta}=\frac{2}{t-t_0(r_\calD)}~~~\text{;}~~~\mathcal{C}_\calD=
\frac{8}{3\left(t-t_0(r_\calD)\right)^2}\,.
\label{LTBmarg2e}
\end{equation}
\end{subequations}
Since the solution was constructed assuming $t>t_0(r)$ for all $r$,  
Eqn. \eqref{LTBmarg2d} immediately shows that $\aDddot<0$ and hence 
acceleration is not possible in this case. Further, 
Eqn. \eqref{LTBmarg2e} shows that
$Q_\calD=\mathcal{C}_\calD-(2/3)\avg{\Theta}^2=0$. Thus the
backreaction term  vanishes for a region described by the marginally
bound LTB solution. This result is not unexpected. We note that
mathematically, the General Relativistic equations \eqref{LTB2a} and
\eqref{LTB2b} describing the evolution of a spherical dust cloud are
identical to the corresponding Newtonian equations. It has been shown
by Buchert, et. al. \cite{buch3} that the backreaction
$\mathcal{Q}_\calD$ in a spherically symmetric \emph{Newtonian} model
of dust, must vanish. Further, we note that \emph{in the marginally
bound case}, the mathematical expressions for the averaged quantities
defined earlier coincide with their corresponding Newtonian
analogues. Hence, for the fully relativistic (marginally bound) case
also, the backreaction must vanish. 

\noindent The spatial Ricci scalar and its spatial average for a general $E(r)$
are given by 
\begin{equation}
^{(3)}\mathcal{R}=\,-\frac{4}{R^2}\left(E+\frac{E^\prime
R}{R^\prime}\right)~~~\text{;}
~~~\avg{\mathcal{R}}=\,-\frac{16\pi}{V_\calD}
\int_0^{r_\calD}{\frac{\frac{\partial}{\partial
r}\left(ER\right)}{\sqrt{1+2E}}~dr} 
\label{LTBmarg3}
\end{equation}
which shows that $^{(3)}\mathcal{R}$ and hence $\avg{\mathcal{R}}$
vanish in the marginally bound case. This is consistent with the
requirement of  Eqn. \eqref{avg12}.

%% file: files/LTBunbd.tex
\section{The unbound LTB solution}
Since the solution with zero spatial curvature fails to produce a
non-trivial backreaction, we consider next the opposite extreme - a
curvature dominated solution in which the contribution to the
Einstein equations due to matter is much smaller than that due to
spatial curvature. Before describing the construction of such a
solution, we present a general treatment of regularity conditions
which an unbound LTB model must satisfy.

%% file: files/regular.tex
\subsection{Regularity conditions on unbound LTB models}
Consider the class of unbound LTB models given by \eqref{LTB4a}. The
functions $M(r)$ and $E(r)$ are to be specified by initial conditions
at $t=t_{in}$, and the choice of scaling $R(r,t_{in})=r$ fixes
$t_0(r)$ as 
\begin{align} 
t_0(r)=t_{in}-\frac{GM(r)}{\left(2E(r)\right)^{3/2}}
\left(\sinh\eta_{in}(r)-\eta_{in}(r)\right)
~~~;~~~\cosh\eta_{in}(r)-1=\frac{2E(r)r}{GM(r)}\,.   
\label{reg1}
\end{align}
The regularity conditions imposed on this model, and their
consequences, are as follows  
\begin{itemize}
\item \underline{{\it No evolution at the symmetry centre}}:\\ 
This is required in order to maintain spherical symmetry about the
same point at all times,  and translates as $\dot R(0,t)=0$ for all
$t$. The right hand side of Eqn. \eqref{LTB2a} must therefore vanish in
the limit $r\to 0$.  
Since the functions involved are non-negative, we assume that we can
write  
\begin{equation}
E(r\to 0)\sim r^\delta~,~\delta >0~~~;~~~
M(r\to 0)\sim r^\alpha~~;~~
R(r\to 0,t)\sim r^\beta f(t)~,~\alpha>\beta\geq 0\,.
\label{reg2}
\end{equation}
Consistency requires $\beta$ to be constant, and our scaling choice
further requires $\beta=1$. We do not require the exponents $\delta$
and $\alpha$ to necessarily be integers.\\ 

\item \underline{{\it No shell-crossing singularities}}:\\
Physically, we demand that an outer shell (labelled by a larger value
of $r$) have a larger area  radius $R$ than an inner shell, at any
time $t$. Unphysical shell-crossing singularities arise when this
condition is not met. Mathematically, this requires 
\begin{equation}
R^\prime(r,t)>0~~~\text{for all $r$, for all $t$.}
\label{reg3}
\end{equation}

\item \underline{{\it Regularity of energy density}}:\\
We demand that the energy density $\rho(r,t)$ remain finite and
strictly positive for all values of $r$ and $t$. Combining this with
Eqns. \eqref{LTB2b} and \eqref{reg3} gives (assuming that $R^\prime$
is finite for all $r$ and since $\beta=1$)
\begin{equation}
\lim_{r\to
0}\rho(r,t)=\,\text{finite}\Rightarrow\,\alpha-1-2\beta=0\Rightarrow\,
\alpha=3\,.   
\label{reg4}
\end{equation}

\item \underline{{\it No trapped shells}}:\\
In order for an expanding shell to not be trapped
initially, it must satisfy the condition $r>2GM(r)$. Near the regular
center, this condition is automatically satisfied independent of the
exact form of $M(r)$, since there $M\sim r^3$.\\
\end{itemize}  
Consider now the function $t_0(r)$ given by \eqref{reg1}. By observing
the behaviour of the functions $(\cosh\eta_{in}-1)$ and
$(\sinh\eta_{in}-\eta_{in})$ for values of $\delta$ equal to, less
than, and greater than $2$, it is easy to check that $t_0(r)$ is finite
at $r=0$ for all values of $\delta$. 
%
However, this involves the assumption that $M(r)$ is positive for
$r\neq 0$. In the limit of $M\to 0$ for all $r$, we find 
\begin{equation}
R\simeq ~\sqrt{2E}\left(t-t_0(r)\right)~~~~;~~~~
t_0(r)\simeq ~t_{in}-\frac{r}{\sqrt{2E}}\,.
\label{reg5}
\end{equation}
Although now, in the limit $r\to 0$, $t_0(r)$ is finite only when
$\delta\leq2$, it will turn out that the integrals involved in
the averaging procedure are insensitive to the behaviour of  $t_0(r)$
in the $r\to 0$ limit, and remain well defined for all positive values
of $\delta$. The expression for $^{(3)}\mathcal{R}$ in
\eqref{LTBmarg3} indicates that the spatial Ricci scalar diverges as
$r\to0$ unless $\delta\geq2$. However, we note that the spatial Ricci
scalar is not a fully covariant quantity and depends on our choice 
of time slicing. The \emph{four}-dimensional Ricci scalar, obtained
after taking the  trace of the Einstein equations as
$^{(4)}\mathcal{R}=8\pi G\rho(r,t)$ is finite at the origin
irrespective of the behaviour of $E(r)$. It is interesting to see how
this cancellation occurs. We have
\begin{equation}
^{(4)}\mathcal{R}=\,^{(3)}\mathcal{R}+2\left\{\left(\frac{\dot
R}{R}\right)^2 +2\frac{\ddot R}{R}\right\}+2\left\{\frac{\ddot
R^\prime}{R^\prime} +2\frac{\ddot R\dot
R^\prime}{RR^\prime}\right\}~~~;~~~
^{(3)}\mathcal{R}=\,-\frac{4}{R^2}\left\{E+\frac{E^\prime
R}{R^\prime}\right\} 
\label{reg6}
\end{equation}
On using the Einstein equation \eqref{LTB2a} we obtain 
\begin{equation}
2\left\{\left(\frac{\dot R}{R}\right)^2+2\frac{\ddot
R}{R}\right\}=4\frac{E}{R^2} ~~~;~~~2\left\{\frac{\ddot
R^\prime}{R^\prime} +2\frac{\ddot R\dot
R^\prime}{RR^\prime}\right\}=2\frac{GM^\prime}{R^2R^\prime}
+4\frac{E^\prime}{RR^\prime} 
\label{reg7}
\end{equation}
which neatly cancels the contribution from $^{(3)}\mathcal{R}$,
leaving precisely $8\pi G\rho(r,t)$ after applying the second Einstein 
equation \eqref{LTB2b}. Hence the $4-$dimensional Ricci scalar does
not impose any further restrictions on the form of $E(r)$. The fact
that the origin is well behaved can also be seen from the  behaviour
of the Kretschmann scalar, given by \cite{hlby}
\begin{equation}
^{(4)}\mathcal{R}_{\mu\nu\sigma\rho}\,
^{(4)}\mathcal{R}^{\mu\nu\sigma\rho}=12\frac{G^2M^{\prime
2}}{R^4 R^{\prime 2}}-32\frac{G^2MM^\prime}{R^5R^\prime}
+48\frac{G^2M^2}{R^6} \,. 
\label{reg8}
\end{equation} 
A condition on the value of $\delta$ \emph{is} obtained, however, by
the regularity of the energy density $\rho(r,t)$, which assumes that
$R^\prime(r,t)$ is not only positive, but also finite for all $r$ and
$t$. Equation \eqref{reg5} shows that unless $\delta=2$, $R^\prime$
either diverges or vanishes at the center, violating this regularity
condition.

%% file: files/latetime.tex
\subsection{Late time solution and curvature dominated unbound models}
The function $R(r,t)$ is an increasing function of time in all the
unbound models described by \eqref{LTB4a}. The Einstein equation
\eqref{LTB2a} then shows that for sufficiently late times $t\gg
t_{in}$, neglecting the term involving $1/R$,  \emph{all} unbound
models have the approximate solution given by \eqref{reg5}.  If on the
other hand, we start with a model which satisfies $GM(r)/(rE(r))\ll1$
for all $r$, then since our scaling assumes that $R=r$ at $t=t_{in}$,
we will have  $GM(r)/(R(r,t)E(r))\ll1$ for all $r$ and for all $t\geq
t_{in}$, and \eqref{reg5} is  then an approximate solution at all
times, the approximation becoming better as $t$ increases. To make
this idea more precise, consider the closed form expression for $t$ in
terms of $R$ obtained by integrating Eqn. \eqref{LTB2a} \cite{tps} 
\begin{equation}
t-t_0(r)=\frac{R^{3/2}}{\left(2GM\right)^{1/2}}\,F\left(\frac{ER}{GM}\right)
~~~;~~~F(x)\equiv\frac{1}{x}(1+x)^{1/2}-\frac{1}{x^{3/2}}\sinh^{-1}
\left(x^{1/2}\right)\,. 
\label{late1}
\end{equation}
Hence, imposing $R(r,t_{in})=r$ we have
\begin{equation}
t_0(r)=t_{in}-\frac{r^{3/2}}{\left(2GM\right)^{1/2}}
F\left(\frac{Er}{GM}\right)\,.  
\label{latet0}
\end{equation}
Let us write $GM(r)/E(r)\equiv\epsilon G\tilde M(r)/E(r)\equiv
\epsilon g(r)$ where $\epsilon$ is a dimensionless positive number
whose value we can control. This relation also defines the functions 
$\tilde M(r)$ and $g(r)$. We can rewrite \eqref{late1} for
$\epsilon\ll1$ as  
\begin{equation}
\sqrt{2E}\left(t-t_0(r)\right)=R\left\{1+\frac{\epsilon
g}{2R}\ln\left(\frac{\epsilon g}{R}\right)-
\frac{\epsilon g}{R}\left(\ln2-\frac{1}{2}\right)+
\mathcal{O}\left(\left(\frac{\epsilon g}{R}\right)^2\right)\right\} \,.
\label{late2}
\end{equation}
Here $\mathcal{O}(x^2)$ represents a power series beginning with a term
of order $x^2$, and we have used a binomial expansion in $\epsilon
g(r)/R$ and the asymptotic expansion for the
inverse hyperbolic sine given by (as $x\to 0$) \cite{math} 
\begin{equation}
\sinh^{-1}\left(\frac{1}{x}\right)=\ln2-\ln x+\frac{1}{4}x^2
+\mathcal{O}\left(x^4\right)\,.
\label{late3}
\end{equation}
The terms in Eqn. \eqref{late2} involving $\epsilon$ vanish as
$\epsilon\to0$, although the expression  in \eqref{late2} cannot be
inverted to get $R=R(r,t)$, due to the presence of the  logarithm. We
can, however, make the following statement. Provided the function
$g(r)/r$ is finite for all values of $r$, then given any  starting
time $t_{in}$, we can choose $\epsilon$ small enough that the terms 
involving $\epsilon$  on the right hand side of Eqn. \eqref{late2} are
all negligible compared to unity. Then, since $R$ increases with time,
these terms will always be 
negligible compared to unity. Alternatively, given some $\epsilon
g/r=GM/(Er)$ which is finite for all $r$, one can always wait for a
sufficiently long time, and find that the $\epsilon$ dependent terms
become smaller compared to unity. In this case, we need not even assume
that $\epsilon$ is small. It is in this sense that the approximation
involved in writing the equations in \eqref{reg5} becomes better  as
$t$ increases (with the caveat that if $\epsilon$ is not small, then
$t_0(r)$ must be given by the full expression \eqref{latet0} and not
the approximation of \eqref{reg5}). This shows that the first of
equations \eqref{reg5} is the correct late time solution 
for all unbound models, with the second being a good approximation
when $\epsilon$ is small. The condition that for some $\epsilon>0$,
$g(r)/r$ be finite for all $r$,  and in particular as $r\to 0$,
implies that $\delta\leq2$ where $\delta$ is defined in
\eqref{reg2}. This is not inconsistent with the requirement $\delta=2$
imposed by the criterion of regularity of energy density.   

Consider now a model which begins
with negligible matter ($\epsilon\to 0$) \emph{and} in which we have
waited for a sufficiently long time ($t\gg t_{in}$). Eliminating
$t_0(r)$ from \eqref{reg5} the approximate solution becomes 
\begin{equation}
R(r,t)=\sqrt{2E(r)}\left(t-\lambda_tt_{in}\right)+\lambda_r r
\label{late4}
\end{equation}
where we have introduced two placeholders $\lambda_r$ and $\lambda_t$ 
which will remind us of the relative magnitudes of various terms. We 
will ultimately set $\lambda_r=\lambda_t=1$. Substituting for $R$ in 
the expression for the domain volume $V_\calD$ in Eqn. \eqref{LTBavg1}, 
we find 
\begin{equation}
V_\calD=(t-\lambda_tt_{in})^3\calI_E+\lambda_r(t-\lambda_tt_{in})^2\calI_{Er} 
+\lambda_r^2(t-\lambda_tt_{in})\calI_{Er^2}+\lambda_r^3\calI_{r^2}
\label{late5}
\end{equation}
where we have defined the domain dependent integrals 
\begin{align}
\calI_E=4\pi\int_0^{r_\calD}{\frac{\sqrt{2E}E^\prime}{\sqrt{1+2E}}~dr}
~~~~&;~~~~
\calI_{Er}=4\pi\int_0^{r_\calD}{\frac{\left(r\cdot2E\right)^\prime}
{\sqrt{1+2E}}~dr}\nonumber\\
\calI_{Er^2}=4\pi\int_0^{r_\calD}{\frac{\left(r^2\cdot\sqrt{2E}\right)^\prime}
{\sqrt{1+2E}}~dr}~~~~&;~~~~\calI_{r^2}=4\pi\int_0^{r_\calD}{\frac{r^2}
{\sqrt{1+2E}}~dr}\,.
\label{late6} 
\end{align}
The sum of the exponents of $\lambda_r$ and $\lambda_t$ in each 
term in \eqref{late5} indicates the relative order of that term  with
respect to the leading $t^3$ term. This approach of treating some 
terms as small compared to others is valid since the various integrals 
which multiply the powers of $t$, are all finite and non-zero. 

We note that the solution in Eqn. \eqref{reg5} actually corresponds to 
Minkowski spacetime, since in this limit the matter content has been
neglected. The corresponding Riemann tensor and Kretschmann scalar are 
exactly zero. The constant time three-spaces are hypersurfaces of
negative curvature, with the three-curvature being determined by the
function $E(r)$. The `FRW' limit of this solution is in fact the
Milne universe; the solution \eqref{reg5} could hence be thought of as
the `Tolman-Bondi' type generalization of the Milne universe. For our 
purpose, it is not a problem that the solution describes Minkowski
spacetime - we know from the initial conditions that
dust matter is present, only its density is negligible compared to the  
curvature term. The form of the solution then allows us to easily
determine if the average scale factor $a_\calD$ undergoes
acceleration. We demonstrate this with explicit examples in the next
subsection. Subsequently, we argue that if a small amount of matter is
introduced, so as to introduce departure from Minkowski spacetime, the
sign of the acceleration of $a_\calD$ is preserved.

%% file: files/accln.tex
\subsection{Condition for late time acceleration}

The expression for the volume $V_\calD(t)$ in \eqref{late5} allows us
to determine the  late time behaviour of the effective scale factor
$\aD(t)\equiv(V_\calD(t)/V_\calD(t_{in}))^{1/3}$. Using a binomial
expansion for $t\gg t_{in}$ in \eqref{late5}, we get
\begin{equation}
3\frac{\aDddot}{\aD}=\frac{\ddot
V_\calD}{V_\calD}-\frac{2}{3}\left(\frac{\dot
V_\calD}{V_\calD}\right)^2
=\frac{2\lambda_r^2}{\calI_Et^4}\left(\calI_{Er^2}-\frac{1}{3\calI_E}
\left(\calI_{Er}\right)^2\right) +\mathcal{O}\left(3\right)
\label{acc1}
\end{equation}
where $\mathcal{O}\left(3\right)$ represents terms involving
$\lambda_r^m\lambda_t^n$, i.e. containing $(1/t^{m+n})$ with
$m+n\geq3$.

We see that the generic late time (i.e. $t\to\infty$) behaviour
of the unbound models under consideration is $\aDddot\to 0$, and that
deviations from zero are small, being  a second order effect. Whether
the approach to $\aDddot=0$ is via an accelerating or decelerating
phase, depends upon the relative magnitudes of the domain integrals
involved. A sufficient condition for an unbound model with negligible
matter to accelerate at late times, is
\begin{equation}
\calP\equiv\calI_{Er^2}-\frac{1}{3\calI_E}\left(\calI_{Er}
\right)^2>0\,.
\label{acc2}
\end{equation}
\begin{figure}[t]
\centering
\includegraphics[width=.65\textwidth]{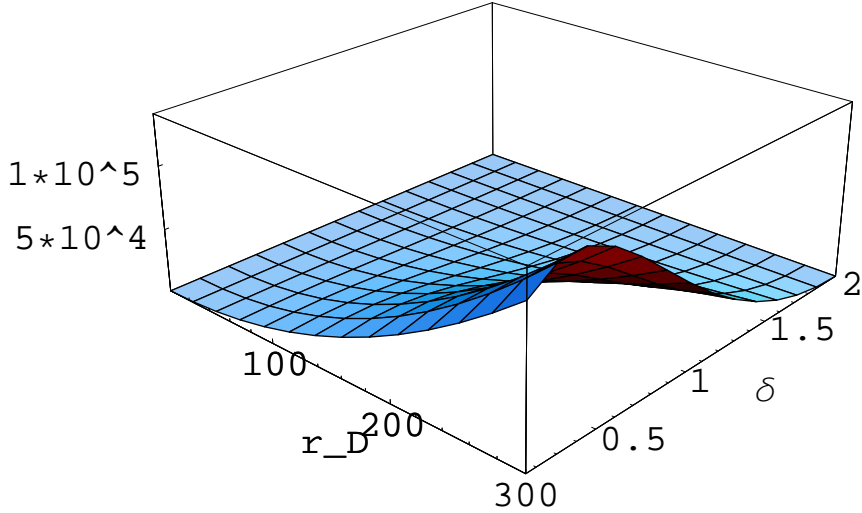}\label{fig1a}
\caption{\small{The function \calP\, defined in the
text, for the power law models described by
$2E(r)=r^\delta$. \calP\, is positive everywhere except along $\delta=2$.}}
\label{fig1}
\end{figure}
To proceed further we need to specify a particular model.As an
explicit example of models admitting acceleration, we consider
the power law models characterized by $2E(r)=r^\delta$, for all $r$,
in some units. (At present we are only demonstrating the existence of
such models, and shall therefore not worry about the physical scales
involved.) Keeping in mind the discussion of Secs. \ref{regular} and
\ref{latetime}, we must strictly speaking only consider the model with
$\delta=2$. The models with $\delta>2$ cannot be considered at all,
since they violate the conditions assumed in Sec. \ref{latetime} which
justified the approximation in Eqn. \eqref{reg5}. The models with
$\delta<2$ on the other hand, contain a Ricci scalar that diverges and
a matter density that vanishes at the center. Despite these
pathologies, we display the results for the models with $\delta<2$ as
an existence proof of acceleration using this very simple
parametrization. Although it is possible to obtain analytical
expressions for the integrals in \eqref{late6} in terms of the
incomplete Beta function, it serves our purpose much better to
numerically evaluate the integrals for various values of $\delta$ and
$r_\calD$, and plot the function \calP~ defined in \eqref{acc2}. The
results are shown in fig.\ref{fig1}. Note that \calP~ vanishes along
the line $\delta=2$, but  is positive \emph{everywhere else} in  the
region plotted, and that the positivity of $\aDddot/\aD$ at late times
is independent of the size of the domain $r_\calD$. We have therefore
obtained a continuous range of parameter values $(\delta,r_\calD)$
which admit late time acceleration.

\begin{figure}[t]
\subfigure[$\calP/\calI_E$]{\includegraphics[width=.45\textwidth]
{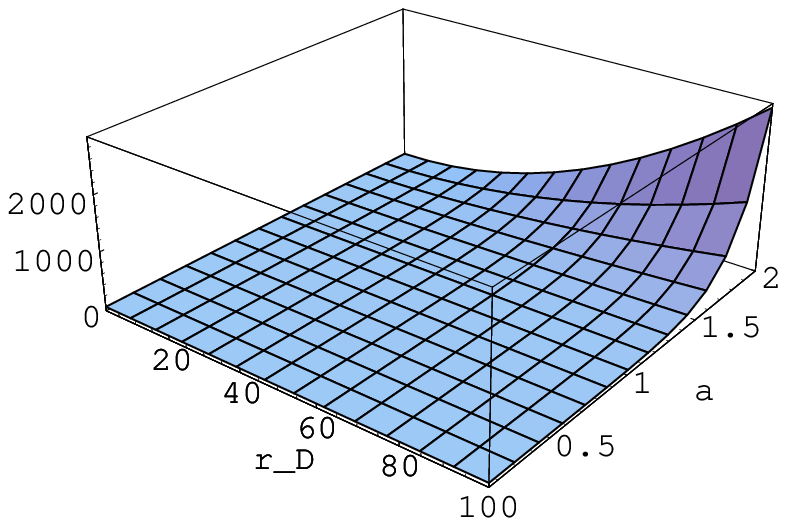}  \label{fig2a}}
\hspace{.05\textwidth}
\subfigure[$\calP/\calI_E$]{\includegraphics[width=.45\textwidth]
{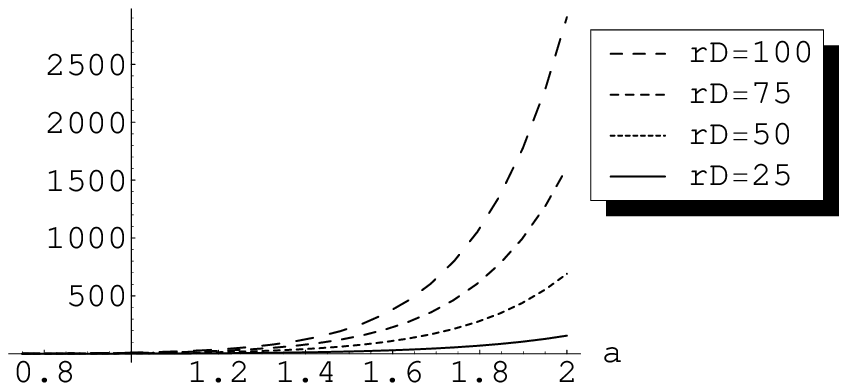}  \label{fig2b}}
\caption{\small{The models described by $2E(r)=r^2/(1+r^a)$. (a) The
scaled function $\calP/\calI_E$. (b) $\calP/\calI_E$ plotted against
$a$ for specific values of $r_\calD$.}}
\label{fig2}
\end{figure}

\begin{figure}
\subfigure[Domain range
$r_\calD=250$]{\includegraphics[width=.45\textwidth]{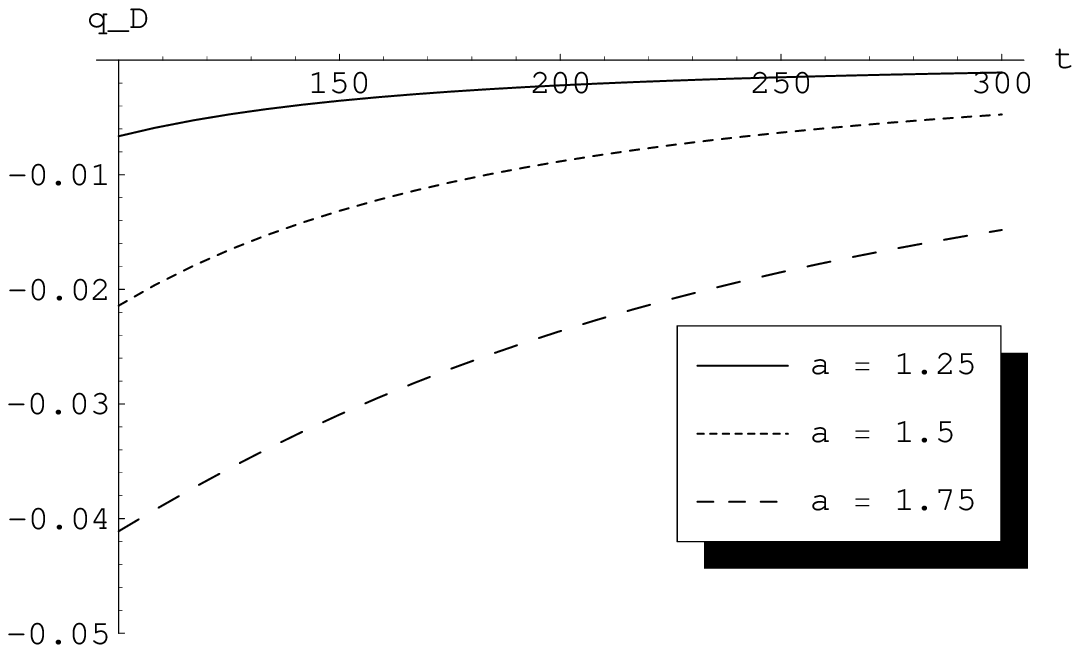}
\label{fig3a}}
\hspace{.05\textwidth}
\subfigure[Exponent $a=1$]{\includegraphics[width=.45\textwidth]{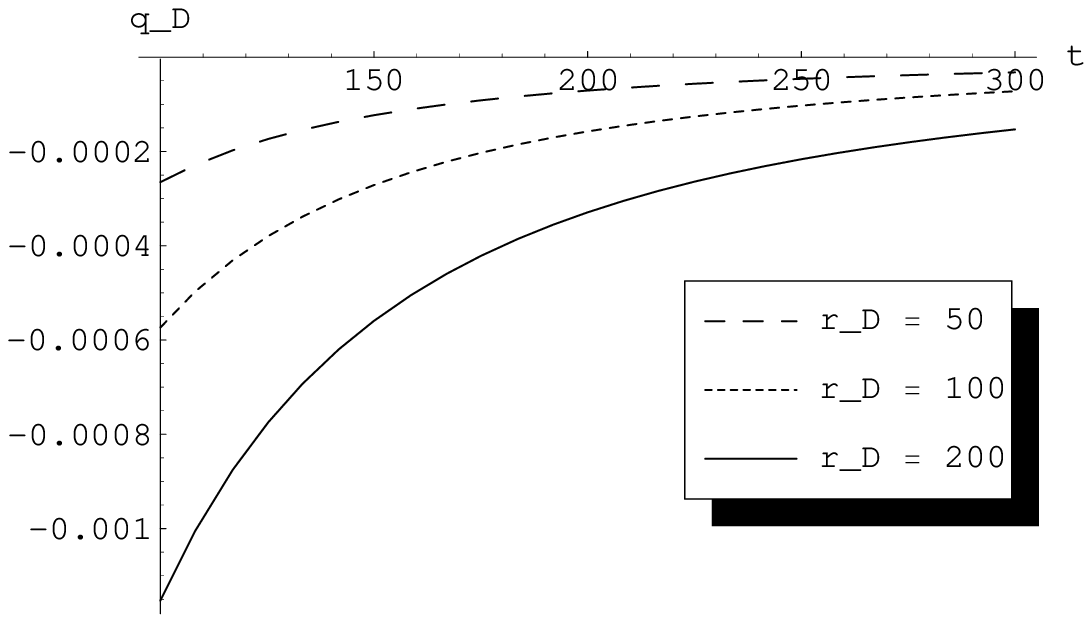}
\label{fig3b}}
\caption{\small{Evolution of $q_\calD$ in the models with
$2E(r)=r^2/(1+r^a)$, plotted for (a) three values of $a$ with
$r_\calD=250$, and (b) three values of $r_\calD$ with $a=1$.}}
\label{fig3}
\end{figure}

In order to demonstrate that the acceleration obtained above is not an
artifact of the singular behaviour of those models, we
construct another set of models which show late time acceleration, and
in which the spatial Ricci scalar $^{(3)}\mathcal{R}$ remains finite
everywhere. Consider the models characterized by the energy function
\begin{equation}
2E(r)=\frac{r^2}{1+r^a}~~~;~~~a>0
\label{acc5}
\end{equation}
where we have again used arbitrary units. Since $a>0$, the $r\to0$
behaviour of these models is $2E\sim r^2$, which satisfies the
regularity conditions of Sec. \ref{regular} and keeps
$^{(3)}\mathcal{R}$ finite at the origin. Also, keeping $a<2$ ensures
the `no shell-crossing' criterion of Sec. \ref{regular}.
The function $\calP/\calI_E$ for these models, which controls the
magnitude of the late time acceleration, is shown in fig. \ref{fig2},
against $r_\calD$ and $a$. For clarity, in the second panel we have
shown $\calP/\calI_E$ against $a$ for specific values of
$r_\calD$. Again, we find that $\calP/\calI_E$ is positive
\emph{everywhere} in the region shown, and hence the models show late
time acceleration for all allowed values of $a$ and $r_\calD$. As an
example, we plot the evolution of the dimensionless quantity $q_\calD$
defined by
\begin{equation}
q_\calD\equiv-\frac{\aDddot\aD}{\aDdot^2}=
2-3\frac{\ddot V_\calD V_\calD}{\left(\dot V_\calD\right)^2}
\label{acc4}
\end{equation}
for various fixed values of $a$ and $r_\calD$.  The results are shown
in fig. \ref{fig3}.  We have used units in which $t_{in}=1$, and have
displayed the evolution for times $t>100\,t_{in}$.
As mentioned earlier, a potentially contentious issue is that in all of
the calculations above, we have actually set $\epsilon=0$. Since the
function $t_0(r)$ approaches its approximation in Eqn. \eqref{reg5} in
a continuous fashion as $\epsilon\to0$, we expect that models with a
non-zero but small matter density will also exhibit the same
qualitative late time behaviour as the ones above. To demonstrate this,
we consider the leading corrections to the function $t_0(r)$ in the
presence of a small but non-zero $\epsilon$. We are still assuming the
late time limit so that the $\epsilon$ dependent terms on the
\emph{right} hand side of Eqn. \eqref{late2} can be neglected (more
precisely, we treat both $\epsilon$ \emph{and} $g/R$ as small
quantities). First, let us rewrite Eqn. \eqref{latet0} as
\begin{equation}
t_0(r)=t_{in}-\frac{h(r)}{\sqrt{2E}}
~~~;~~~h(r)\equiv
r\left(\frac{Er}{GM}\right)^{1/2}\,F\left(\frac{Er}{GM}\right)=
r\left\{1+\frac{\epsilon g}{2r}\ln\left(\frac{\epsilon g}{r}\right)
+\mathcal{O}\left(\frac{\epsilon g}{r}\right)\right\}\,.
\label{acc6}
\end{equation}
It is easy to check that in the late time limit, the expression for
volume $V_\calD$ becomes
\begin{equation}
V_\calD=(t-\lambda_tt_{in})^3\calI_E +
\lambda_r(t-\lambda_tt_{in})^2\calI_{Eh}
+\lambda_r^2(t-\lambda_tt_{in})\calI_{Eh^2} + \lambda_r^3\calI_{h^2}
\label{acc7}
\end{equation}
where $\calI_E$ is the same as defined in \eqref{late6}, and the
remaining integrals are defined analogously to those in
\eqref{late6},
\begin{equation}
\calI_{Eh}=4\pi\int_0^{r_\calD}{\frac{(2Eh)^\prime}
{\sqrt{1+2E}}~dr} ~~~;~~~
\calI_{Eh^2}=
4\pi\int_0^{r_\calD}{\frac{(h^2\sqrt{2E})^\prime}
{\sqrt{1+2E}}~dr}
~~~;~~~\calI_{h^2}=4\pi\int_0^{r_\calD}{\frac{h^2h^\prime}
{\sqrt{1+2E}}~dr}  \,.
\label{acc8}
\end{equation}
In arriving at Eqn. \eqref{acc7}, we have neglected terms involving
$(g/R)(\eplog)$, $\epsilon(g/R)\ln(g/R)$ and terms of order
$\mathcal{O}(\epsilon g/R)$ on the right hand side of
Eqn. \eqref{late2}. In order to proceed as before, we further assume
that these leading order corrections are \emph{smaller} than the terms
of order $\lambda_r^2$ coming from the binomial expansion of $V_\calD$
in Eqn. \eqref{acc7}. This is essential in order to be able to make a
statement analogous to \eqref{acc2}, and can be ensured by choosing
$\epsilon$ small enough, without setting it exactly to zero. The
condition for late time acceleration in this situation becomes
\begin{equation}
\calP_h\equiv\calI_{Eh^2}-\frac{1}{3\calI_E}\left(\calI_{Eh}
\right)^2>0\,.
\label{acc9}
\end{equation}
For $\epsilon<e^{-1}$, we have $0<\epsilon<-\eplog<1$, and the leading order
terms in the expansion of $h(r)$ contain $(\eplog)$ (assuming that the
$r$ dependent coefficients are well behaved for all $r$). On expanding the
integrals in \eqref{acc9} to this leading order, we find for the
function $\calP_h$,
\begin{align}
\calP_h=\calP +(-\eplog)\calP^{(\eplog)}
+\mathcal{O}\left(\epsilon,(\eplog)^2\right) ~~~&;~~~
\calP^{(\eplog)}\equiv\frac{2}{3\calI_E}\calI_{Er}\calI_{Er}^{(\eplog)}-
\calI_{Er^2}^{(\eplog)}\nonumber\\
\calI_{Er}^{(\eplog)}= 4\pi\int_0^{r_\calD}{\frac{G\tilde M^\prime}
{\sqrt{1+2E}}~dr} ~~~&;~~~
\calI_{Er^2}^{(\eplog)}= 4\pi\int_0^{r_\calD}{\frac{(2G\tilde
M r/\sqrt{2E})^\prime}{\sqrt{1+2E}}~dr}
\label{acc10}
\end{align}
\noindent where $\tilde M$ is defined by $M=\epsilon\tilde M$, \calP\, is
defined in \eqref{acc2} and the integrals
$\calI_{Er}^{(\eplog)}$ and $\calI_{Er^2}^{(\eplog)}$ give the leading
order corrections to $\calI_{Er}$ and $\calI_{Er^2}$
respectively. The function $\calP_h$ of \eqref{acc9} replaces $\calP$
in Eqn. \eqref{acc1}. This shows that a non-zero $\epsilon$ brings in
an additional correction to $\aDddot/\aD$ which is of order
$\lambda_r^2(\eplog)$. We have already neglected terms of order
$(g/R)(\eplog)$, and since $(g/R)$ is small \emph{because} $t$ is
large, we should therefore also ignore terms of order
$\lambda_r(\eplog)$, $\lambda_t(\eplog)$ and higher. Hence the
correction given by $\calP^{(\eplog)}$ (provided it is finite), should
be ignored. We therefore see explicitly that within the late time
approximation, we can always have a non-zero but small enough
$\epsilon$ which does not affect the sign of the acceleration at the
leading order.

A rigorous argument to demonstrate acceleration in models with
non-zero matter in general would, of course, require a complete
numerical evolution of the LTB equations, and this work is in
progress.  Our semi-numerical analysis, however, throws open the
possibility of a large class of models which may show late time
acceleration after averaging.

Our results for the zero matter limit also
point to the need for caution in interpreting  acceleration  -  by 
suitable choices of the energy function $E(r)$ one could obtain
acceleration, even though in this limit the spacetime coincides with
Minkowski spacetime. A similar demonstration was earlier given by
Ishibashi and Wald \cite{wald}. They showed that by suitably joining
two negative curvature slices (hyperboloids) in Minkowski spacetime
one can construct a spatial region which exhibits acceleration. In
contrast though, the slicing we choose is physically motivated, so as
to coincide with the FRW slicing when matter is included. Our own
interest of course  was in  demonstrating, by adding matter beyond the
Minkowski limit, that acceleration is possible in a single expanding
LTB region. 

%% file: files/anlytc.tex
\subsection{An analytical example ($r=0$ excluded)}
In this subsection we will follow a slightly different approach
and try to construct an accelerating model from purely analytic
arguments. We begin with a domain in which $t_0(r)>0$ for all $r$.
We now use the approximate solution \eqref{reg5} in the expression
for volume in \eqref{LTBavg1} and keep the integration limits
unspecified as $r_{1}$, $r_{2}$, i.e. 
$V_\calD=4\pi\int_{r_1}^{r_2}{R^\prime R^2/\sqrt{1+2E}dr}$.
At late times $t$, by treating $t_0(r)$ in its entirety as a small
quantity compared to $t$, we find
\begin{equation}
\aDddot\propto\frac{1}{\calI_1t^4}\left(\calI_1\calI_3
-\frac{\calI_2^2}{3}\right)
\label{anlytc1}
\end{equation}
where the constant of proportionality is positive, $\epsilon$
dependent terms have been neglected and we have defined
the integrals\footnote{Factors of $4\pi$ have been absorbed in the
proportionality constant.}
\begin{equation}
\calI_1=\frac{1}{3}\int_{r_1}^{r_2}{\frac{\left((2E)^{3/2}
\right)^\prime}{\sqrt{1+2E}}~dr}~~~;~~~
\calI_2=\,-\int_{r_1}^{r_2}{\frac{\left((2E)^{3/2}t_0
\right)^\prime} {\sqrt{1+2E}}~dr}~~~;~~~
\calI_3=\int_{r_1}^{r_2}{\frac{\left((2E)^{3/2}t_0^2 \right)^\prime} 
{\sqrt{1+2E}}~dr}\,.
\label{anlytc2}
\end{equation}
Note that $\aDddot$ is still a second order quantity at late
times. The integral $\calI_1$, which is essentially the same as
$\calI_E$ defined in \eqref{late6}, is of the form
$\int{x^{1/2}/\sqrt{1+x}dx}$. This can be evaluated exactly and
$\calI_1$ is positive provided $E(r)$ is an increasing function of 
$r$, which we henceforth assume. Let us also assume that $t_0(r)$ is
an increasing function of $r$, this requires $E(r)$ to increase faster  
than $r^2$. Although this is not consistent with our arguments of
Sec. \ref{latetime}, notice that simultaneously requiring $t_0>0$ and 
$t_0^\prime>0$ places a restriction on the minimum value that $r$ can 
take. For example, if $2E=(r/r_0)^m$, $m>2$ then these conditions
imply (using Eqn. \eqref{reg5}) that $r\geq
r_1>(r_0^m/t_{in}^2)^{1/(m-1)}$. Since the origin is necessarily
excluded, the function $GM/(Er)$ will remain finite for all allowed
values of $r$, even though $E(r)$ rises faster than $r^2$. So the
assumptions on which the later arguments were built, are not violated.

The condition for positivity
of \aDddot\, at late times is now $\tilde\calP\equiv\calI_1\calI_3
-\calI_2^2/3>0$. Our assumptions above ensure that the integrands of
$\calI_1$, $\calI_2$ and $\calI_3$ are all positive. We denote
\begin{equation}
f\equiv(2E)^{3/2}~~~;~~~p\equiv\frac{t_0}{t_{in}}=
1-\frac{r}{t_{in}(2E)^{1/2}}\,.  
\label{anlytc3}
\end{equation}
The positivity condition can be written as
\begin{equation}
\left\{\int_{r_1}^{r_2}{\frac{f^\prime}{3\sqrt{1+2E}}~dr}
\right\}
\left\{\int_{r_1}^{r_2}{\frac{t_{in}^2(fp^2)^\prime}{\sqrt{1+2E}}~dr}    
\right\}>
\left\{\int_{r_1}^{r_2}{\frac{t_{in}(fp)^\prime}{\sqrt{3}\sqrt{1+2E}}~dr}  
\right\}^2   \,.
\label{anlytc4}
\end{equation}
A set of sufficient conditions for this relation to hold, is
\begin{equation}
f^\prime>(fp)^\prime~~~;~~~(fp^2)^\prime>(fp)^\prime\,.
\label{anlytc5}
\end{equation}
Some straight-forward algebra reduces these conditions to
\begin{equation}
\frac{f^\prime}{f}>\frac{p^\prime}{1-p}~~~\textrm{and}~~~
\frac{f^\prime}{f}>\frac{p^\prime}{1-p}\frac{2p-1}{p}   
\label{anlytc6}
\end{equation}
respectively. By definition, $0<p<1$ since we require $t_0>0$ for all
$r$. This implies $(2p-1)/p<1$ for all $r$, so if the first relation
in \eqref{anlytc6} holds, then so does the second. The first relation,
in terms of $E(r)$, reduces to $E^\prime/E>(-1/r)$ which is
necessarily true. Hence we have obtained a class of solutions which
show late time acceleration of the effective scale factor \aD, with
the caveat that we must exclude a sphere around the origin from our
domain of integration, the radius of this sphere being determined by
the form of $E(r)$.

%% file: files/discuss.tex
\section{Discussion}
 We have shown that it is possible within the framework of classical
 General Relativity, to construct models of universes in which the
 average behaviour of spatial slices is that of accelerating
 expansion. Although the LTB models are unrealistic (since they place
 us at the center of the Universe), they are useful in building
 intuition. Especially, since LTB is an exact solution, it helps
 towards a deeper understanding of averaged inhomogeneous cosmological 
 solutions of Einstein equations. In particular, our solution could be
 assumed to apply not necessarily to the whole Universe, but only to a
 local underdense region such as a void. The volume average of the
 late Universe is dominated by voids, and structures occupy a tiny
 fraction of the volume. The average over such a distribution does not
 lead to a FRW model. The curvature of voids can be estimated to be
 proportional to minus the square of their Hubble expansion rate and
 thus must be negative \cite{Rasanen2}. The negative curvature LTB
 model discussed in this paper could be useful for describing such a
 situation. 

 In both the examples which we gave, namely the
 power law  models and the models given by Eqn. \eqref{acc5}, the
 qualitative behaviour of the evolution of the effective scale  factor
 was independent of the size of the averaging domain. Whereas the
 power law models were pathological in the sense that,
 among other things, their spatial Ricci scalars diverged at the
 symmetry center, the second class of models  described by
 \eqref{acc5} had no such pathology.

 The point to be emphasized, though, is that our analysis
 clearly shows the importance of curvature (expressed as a non-zero
 energy function $E(r)$) in causing the late time acceleration. The
 fact that the limit of vanishing matter density ($\epsilon\to0$) is
 well defined, means that the qualitative behaviour of curvature
 dominated models is not expected to change by adding a finite but
 small amount of matter.

 A few remarks comparing the results of the present paper with the
 concordance model in standard cosmology are in order. One could
 assert that observational data show that our currently accelerating
 universe has zero spatial curvature. On the other hand, our LTB model
 with zero spatial curvature (the marginally bound case) shows no
 acceleration. It might hence appear that our curvature dominated LTB
 model is of very limited interest. However, observations in the late
 Universe must not be matched only with a zero-curvature
 Universe. This assumption may be good in the early stages of
 evolution, but it is just a fitting ansatz to the late-time
 inhomogeneous Universe. What the consideration of averaged 
 inhomogeneities in the present and other similar papers demonstrates 
 is that the real Universe could in principle have negative curvature
 and yet exhibit acceleration. If this were indeed to be the case, it
 could eliminate the need for a dark energy. 

It might also be said that observational data show that the Universe
has up to thirty percent of its content in luminous and dark matter
and hence our low density curvature dominated model showing
acceleration does not meet this criterion. However, in reality our
model is in disagreement with the concordance model, which while being
one of the simplest and most favoured possibilities, need not turn out
to be the final answer. The consideration of averaged inhomogeneities
shows that low density, curvature dominated models can also in
principle fit the observational data. This issue should hence be
regarded as an open one. Furthermore, as noted above, our LTB model
could by itself be considered relevant for describing a locally
underdense region such as a void.

Our results highlight the intimate connection between the averaged
spatial curvature, and the evolution of the kinematic back-reaction,
as anticipated from the integrability condition given by
Eqn. \eqref{avg12}. It will be interesting to consider the more
general dust models described by the metric \eqref{avg1}, and enquire if
acceleration can again be produced in the approximation where the
averaged three-curvature dominates over the averaged matter density
during some epoch of cosmic evolution.

\section*{Acknowledgements}

 We thank Sarang Sane and Rakesh Tibrewala for useful discussions, and
 Thomas Buchert and Syksy R\"as\"anen for comments on an earlier
 version.